\documentclass[fleqn,usenatbib]{mnras}

\usepackage{newtxtext,newtxmath}
\usepackage{braket}
\usepackage{soul}
\usepackage[T1]{fontenc}

\DeclareRobustCommand{\VAN}[3]{#2}
\let\VANthebibliography\thebibliography
\def\thebibliography{\DeclareRobustCommand{\VAN}[3]{##3}\VANthebibliography}

\usepackage{graphicx}	% Including figure files
\usepackage{amsmath}	% Advanced maths commands

\title[Dynamics in mAGN]{The formation of mini-AGN disks around IMBHs and their dynamical implications}

\author[Rozner et al.]{
Mor Rozner,$^{1,2,3,4}$\thanks{E-mail: 
morozner@ias.edu}
Alessandro A. Trani,$^{5}$
Johan Samsing$^{5}$
and Hagai B. Perets$^{2,6}$
\\
$^{1}$Institute of Astronomy, University of Cambridge, Madingley Road, Cambridge CB3 0HA, UK\\
$^{2}$Technion - Israel Institute of Technology, Haifa, 3200002, Israel\\
$^{3}$Gonville \& Caius College, Trinity Street, Cambridge, CB2 1TA, UK\\
$^{4}$Institute for Advanced Study, Einstein Drive, Princeton, NJ 08540, USA\\
$^{5}$Niels Bohr International Academy, Niels Bohr Institute, Copenhagen, Denmark\\
$^{6}$ Astrophysics Research Center of the Open University (ARCO), The Open University of Israel, P.O. Box 808, Raa'nana 4353701, Israel
}

\date{Accepted XXX. Received YYY; in original form ZZZ}

\pubyear{\the\year{}}

\begin{document}
\label{firstpage}
\pagerange{\pageref{firstpage}--\pageref{lastpage}}
\maketitle

\begin{abstract}
This study explores the formation and implications of mini-active galactic nuclei (mAGN) disks around intermediate-mass black holes (IMBHs) embedded in gas-rich globular/nuclear clusters (GCs). We examine the parameter space for stable mAGN disks, considering the influence of IMBH mass, disk radius, and gas density on disk stability. The dynamics of stars and black holes within the mAGN disk are modeled, with a focus on gas-induced migration and gas dynamical friction. These dynamical processes can lead to several potentially observable phenomena, including the enhancement of gravitational wave mergers (particularly IMRIs and EMRIs), and the occurrence of mili/centi-tidal disruption events (mTDEs/cTDEs) with unique observational signatures. We find that gas hardening can significantly accelerate the inspiral of binaries within the disk, potentially leading to a frequency shift in the emitted gravitational waves. Additionally, we explore the possibility of forming accreting IMBH systems from captured binaries within the mAGN disk, potentially resulting in the formation of ultraluminous X-ray sources (ULXs). The observational implications of such accreting systems, including X-ray emission, optical signatures, and transient phenomena, are discussed. Furthermore, we investigate the possibility of large-scale jets emanating from gas-embedded IMBHs in GCs. While several caveats and uncertainties exist, our work highlights the potential for mAGN disks to provide unique insights into IMBH demographics, accretion physics, and the dynamics of GCs.
\end{abstract}


%%%%%%%%%%%%%%%%%%%%%%%%%%%%%%%%%%%%%%%%%%%%%%%%%%

%%%%%%%%%%%%%%%%% BODY OF PAPER %%%%%%%%%%%%%%%%%%
\section{Introduction}
Intermediate-mass black holes (IMBHs), with masses of $10^2-10^5 \ M_\odot$, occupy a unique range in the black hole mass spectrum. While stellar-mass black holes (SBHs) and supermassive black holes (SMBHs) are well-established through numerous observations, IMBHs have remained elusive despite extensive searches.

However, in recent years there has been a surge in evidence supporting their existence. The detection of GW190521, a gravitational wave signal potentially arising from the merger of two SBHs, hints at the formation of an IMBH with a mass of approximately $142 \ M_\odot$ \citep{Abbott2020_IMBH}. Prior to this, the most compelling evidence for IMBHs stemmed from observations of active galactic nuclei (AGNs), like the one in the galaxy NGC4395, hosting a black hole with a mass of $3\times 10^5 \ M_\odot$ solar masses \citep{Peterson2005}. One of the candidate GCs to host an IMBH is $\omega$-Cen, which has evidence for a stellar disk, suggesting the past existence of an inner gaseous disk \citep{VandeVen2006,MastrobuonoBattisti2013}.

Several formation channels have been proposed for IMBHs, including a direct collapse of gas at high redshifts \citep[e.g.][]{EardleyPress1975,Begelman2006}; remnants of the early pop III ultra-massive stars in the Universe \citep[e.g.][]{MadauRees2001}; or runaway collisions and mergers of massive stars or black holes with main-sequence stars \citep[e.g.][]{Ebisuzaki2001,PortegiesZwart2002,Fujii2024,Rose2021}.

IMBHs are expected to play a crucial role in various astrophysical phenomena: They may serve as seeds for the growth of SMBHs at the centers of galaxies \citep[e.g.][]{MadauRees2001, Silk2017};  mergers of IMBHs with other compact objects could generate detectable gravitational waves \citep[e.g.][]{AmaroSeoane2007,Mandel2008,Gualandris2009,FragioneLeigh2018,FragioneTidalGW2018}; and tidal disruption events (TDEs) may occur when stars venture too close to IMBHs \citep[e.g.][]{MacLeod2016,FragioneTidalGW2018}.

The growing evidence for IMBHs, together with their potential astrophysical significance, has spurred renewed interest in their formation, evolution, and observational signatures.

It has been hypothesized that IMBHs could form and reside within globular clusters (GCs). Due to the processes of mass segregation and dynamical friction, these IMBHs are anticipated to migrate towards the central regions of GCs. Within these dense environments, they are likely to form binaries with stars or stellar-mass black holes \citep{SigurdssonHernquist1993}. These binaries, through a process of hardening, could eventually become sources of gravitational waves (GWs) with distinctive mass ratios, namely extreme mass ratios (EMRIs) with $10^{-8}\lesssim q=m_2/m_1 \lesssim 10^{-5}$, and intermediate-mass ratios (IMRIs) with $10^{-5}\lesssim q\lesssim 10^{-2}$ \citep{AmaroSeoane2007}. Future space-based GW detectors like LISA may be able to observe these unique signals \citep{AmaroSeoane2007,AmaroSeoane2022}. 

Historically, GCs were believed to host a simple stellar population arising from a single burst of star formation. However, it is now understood that the majority of GCs contain at least two distinct populations (e.g. \citep{BastianLardo2015}). While the origin of these multiple populations remains unclear, any episode of star formation necessitates the presence of a substantial amount of gas. Consequently, the first population stars and compact objects formed in GCs would later be embedded in a gas-rich environment that fostered the subsequent formation of second population stars. This gaseous environment could significantly influence the dynamical evolution of stars within the cluster.

Recent research has demonstrated that such gas-rich environments can serve as fertile grounds for various astrophysical phenomena, including gas-induced binary formation \citep{Rozner2023,Rowan2023,DodiciTremaine2024,Whitehead2024a} and the generation of gravitational waves \citep{RoznerPerets2022}. Other potential effects encompass the accretion-driven growth of stars and compact objects, gas-induced hierarchical mergers, spin alignment, and enhanced mass segregation and/or contraction of the cluster \citep[see][and references therein]{Leigh2013,Lei+14,Perets2022}. Recently, many studies dealt with various aspects of the uniqueness of binary evolution in gas-rich environments (e.g. \citealp{LiLai2022, LiLai2023a,LiLai2023b,Mishra2024,
Rowan2024,ONeill2024,
Whitehead2024b} and references therein), and still our physical understanding of these systems is far from being complete.

In this work, we study the evolution of IMBHs in gas-rich GCs.
In this gas-rich environment, the IMBH would become embedded in gas, leading to the formation of a circumstellar disk analogous to those observed around supermassive black holes in active galactic nuclei (AGNs). We refer to such a system as a mini-AGN (mAGN). Notably, the formation of an mAGN does not require second-generation star formation necessarily; an accretion of gas from the interstellar medium or other external sources onto the cluster could also trigger this phenomenon \citep{LinMurray2007,Naiman2011}.

We study the conditions necessary for the formation of an mAGN disk and investigate its potential observational and theoretical implications for the dynamical evolution of stars within the cluster. The physics governing stars and compact objects embedded in AGN disks is complex and has been the subject of extensive research, particularly in recent years \citep[e.g.][]{Ostriker1983, Artymowicz1993, McKernan2012, Stone2017, CantielloLin2021, Tagawa2020, Li2023, Whitehead2023,DodiciTremaine2024}. These studies have revealed a rich tapestry of dynamical processes, including gas-induced mergers of stars and stellar-mass black holes, migration within the disk, the capture of stars and compact objects into the disk, accretion-driven growth, and the production of unique explosive transients and gravitational wave sources \citep[e.g.][]{Artymowicz1993, McKernan2012, Bartos2017, Stone2017, Tagawa2020, Grishin2021Sne, Samsing2022, vaccaro2023}. It is reasonable to expect that many of these processes (and some others), albeit with modifications, could also transpire within the context of mAGNs.

In section \ref{sec:parameter_space}, we characterize the available parameter space for the mAGNs. In section \ref{sec:motion in gas} we discuss different models for motion in gas. In subsection \ref{subsec:disk-migration}, we discuss disk migration. In subsection \ref{sec:gas_dynamical_friction}, we briefly introduce gas dynamical friction. 
In section \ref{sec:dynamical_processes}, we discuss some of the possible dynamical processes expected in mAGNs, including alignment into a mAGN (subsec. \ref{subsec:alignment}), GW mergers (subsec. \ref{subsec:GW_rate}) and TDEs (subsec. \ref{subsec:TDE}). In section \ref{sec:accreting_IMBH} we discuss the formation of accreting IMBHs from capture binaries. In section \ref{sec:jets} we discuss the formation of large-scale jets. We then discuss our results and possible implications in section \ref{sec:discussion}, introduce possible caveats in \ref{sec:caveats} and finally conclude and summarize in \ref{sec:summary}. 

\section{Parameter space of the disk} \label{sec:parameter_space}

Let us consider a globular cluster during the formation of its second (or further) generation of stars, which also harbors an IMBH. During this phase, a pre-existing IMBH within the GC would be embedded in gas, leading to the formation of an accretion disk around the IMBH. We aim to constrain the parameters of this disk using analytical arguments and a numerical solution of the accretion disk equations as presented in \citet{sirko2003,Gangardt2024}.

A mass inflow into the disk persists as long as gas is present within the cluster. However, the gas is expelled from the cluster upon the emergence of the first supernovae, establishing a maximum typical gas lifetime of roughly $\sim 10-10^2 \ \rm{Myr}$ \citep{BastianLardo2015}. It should be noted that this assumption is somewhat conservative. The continuous mass loss from giant stars provides a sustained gas supply, a portion of which is expected to accrete onto the IMBH even at later times. Furthermore, as previously mentioned, additional potential sources of gas accretion onto the cluster could replenish the gas supply to the IMBH. 

Afterward, the gas depletes within a typical viscous time,  given by  

\begin{align}
&\tau_v=\frac{1}{\alpha \Omega}\left(\frac{h}{r}\right)^{-2}= \\
\nonumber
&= \left(\frac{0.01}{\alpha}\right)\left(\frac{6.47 \times 10^{-10} \ \rm{sec}^{-1}}{\Omega}\right)\left(\frac{h/r}{0.01}\right)^{-2}\approx 
49 \ \rm{Myr}
\end{align}
\noindent
where $\alpha$ is the Shakura-Sunyaev parameter \citep{ShakuraSunyaev1973}, $\Omega$ is the Keplerian angular frequency and $h/r$ is the aspect-ratio of the disk. 

\begin{figure*}
    \includegraphics[width=\textwidth]{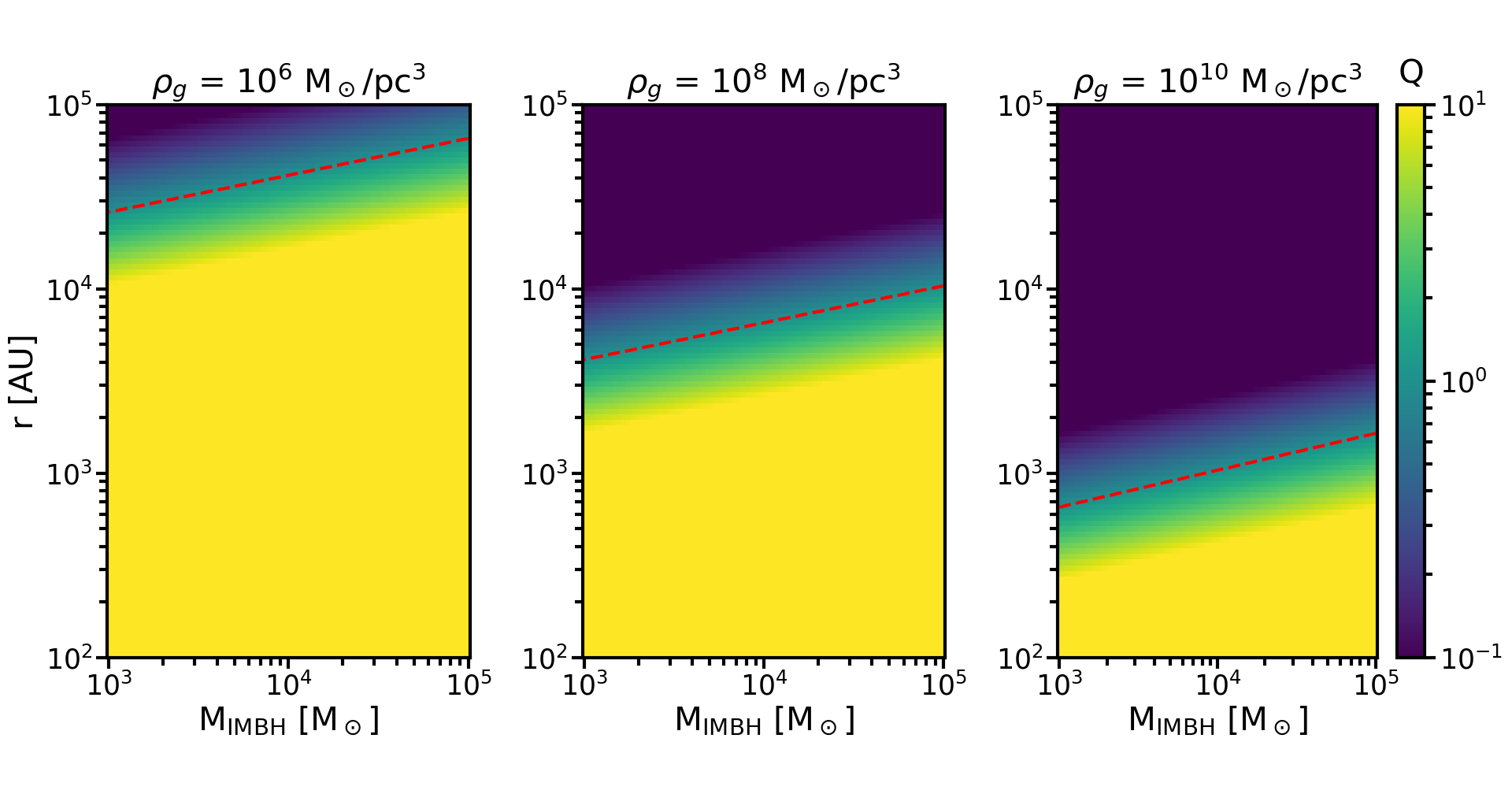}
    \caption{ The Safronov-Toomre parameter (eq. \ref{eq:Q} as calculated for different possible disk parameters, assuming a temperature of $T=10^5 \ \rm{K}$. The red dashed lines indicate the maximal allowed radius ($Q=1$), such that all of the radii below the dashed line will have a stable disk ($Q>1$), and all the radii above the dashed line we will have an unstable disk ($Q<1$). Different panels stand for different densities, assuming a constant density through the disk. 
    }
    \label{fig:Q}
\end{figure*}

We calculate the Safronov-Toomre stability parameter \citep{Safronov1960,Toomre1964} for different disk radii and IMBH masses.  
The gravitational stability of the disk could be determined by 

\begin{equation}\label{eq:Q}
Q = \frac{c_{s,mAGN}\Omega}{\pi G \Sigma}= \frac{h M_{\rm IMBH}}{\sqrt{2}\pi \Sigma R^2}=\frac{M_{\rm IMBH}}{\sqrt{8}\pi \rho_{\rm mAGN}R^3}
\end{equation}
\noindent
where $c_{\rm{s,mAGN}}$ is the sound speed in the mAGN, $\Sigma$ is the surface density, $h$ is the scale height of the mAGN disk, $M_{\rm IMBH}$ is the mass of the IMBH, $\rho_{mAGN}$ is the midplane density  and $R$ is the distance from the center. Note that we relate here everywhere to the midplane gas density.
In Fig. \ref{fig:Q}, we present the value of $Q$ for different combinations in the parameter space. 
The yellow area marks the available parameters for a stable disk according to this criterion, i.e. $Q\geq 1$. The Safronov-Toomre parameter shown here is local in the sense that we compare the local contribution from gravity to the one from thermal pressure.

\begin{figure*}%[ht]
	\includegraphics[width=1\textwidth]{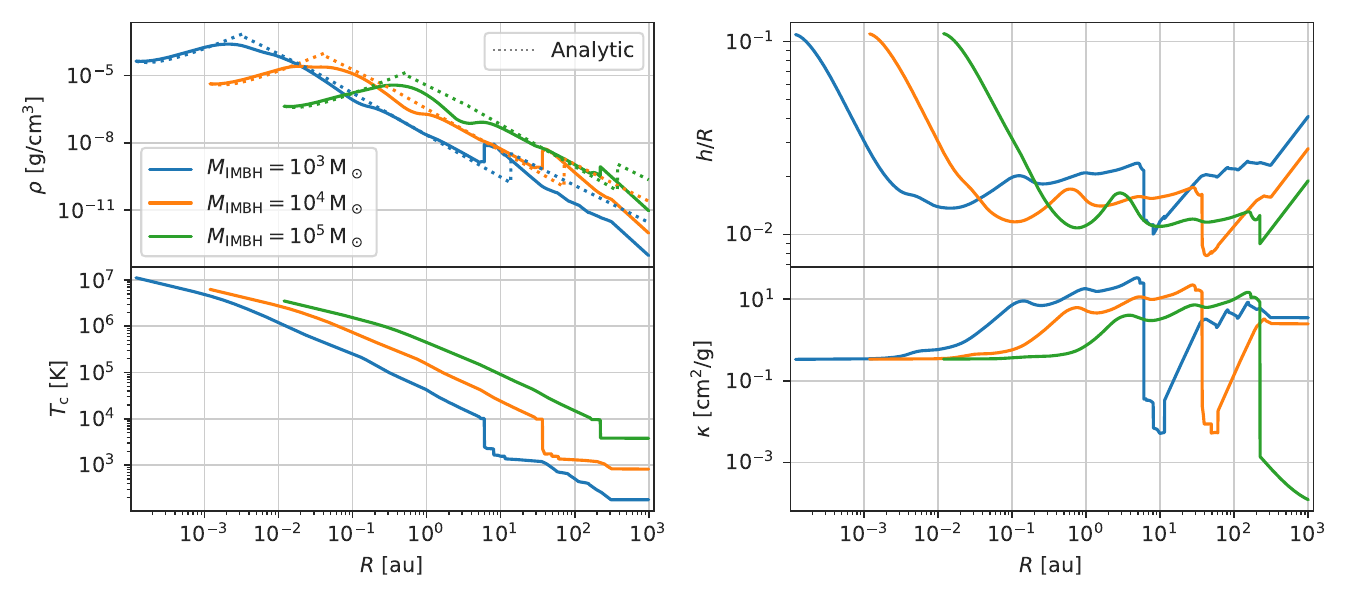}
	\caption{Various profiles of the three \citet{sirko2003} disks around an IMBH considered here. Top-left panel: gas density. Top-right panel: aspect ratio of the disk. Bottom-left panel: midplane temperature. Bottom-right panel: opacity. 
		Blue, orange and green lines indicate an IMBH mass of $M_{\rm IMBH} = 10^3, 10^4$ and $10^5 \, M_\odot$, respectively.}
	\label{fig:Sirko}
\end{figure*}

In Figure \ref{fig:Q}, we present the Safronov-Toomre parameter for different choices of $M_{\rm IMBH}$, radii, and gas densities, as calculated using \ref{eq:Q}. We set the available parameter space for stable mAGN disks and specifically the largest radius allowed for a stable disk, given a mass of a central IMBH. The typical radius of a mAGN disk of gas density $10^{10} \ M_\odot \ \rm{pc}^{-3}$ surrounding a $10^4 \ M_\odot$ IMBH will be $\sim 10^3 \ \rm{AU}.$ 

In Figure \ref{fig:Sirko}, we present the numerical solution for a mAGN using the \citet{sirko2003} equations. We modeled three $\alpha$-disks (i.e. with the viscosity being proportional to the total pressure, see \citealt{Haiman2009}), for IMBH masses $M_{\rm IMBH} = 10^3, 10^4,$ and $10^5 \, M_\odot$, assuming a viscosity parameter of $\alpha=0.1$ and a mass accretion $\dot M = 0.5  \, {\dot M}_{\rm Edd}$, where 
\begin{equation}
{\dot M}_{\rm Edd}  =  \frac{L_{\rm Edd}  }{\epsilon c^2 }\,,
\end{equation}
is the Eddington accretion rate, $L_{\rm Edd}$ is the Eddington luminosity and $\epsilon=0.1$ is the radiation efficiency (for more details see \citealp{Gangardt2024}). 

The models with $M_{\rm IMBH} = 10^3$ and $10^4 \, M_\odot$ exhibit the so-called opacity gap, the drop in opacity at temperatures of $T \approx 10^3$--$10^4 \rm\, K$, at a radius of $10-70 \,\rm{AU}$, while the $M_{\rm IMBH} = 10^5 \, M_\odot$ disk, being more massive and hotter, enters the opacity gap only in the outermost region (${\sim}200 \ \rm{AU}$).

The gas density in the disk is higher than in AGN disks, as the gas is concentrated in a smaller region, as expected also from the crude estimate (see also \citealp{CantielloLin2021})
\begin{align}
\rho_g \sim \frac{M_{\rm IMBH}}{\sqrt{8}\pi R^3}\sim  
10^{10} \frac{M_\odot}{\rm{pc}^3} \frac{M_{\rm IMBH}}{10^{4}M_\odot }\left(\frac{10^3 \ \rm{AU}}{R_{\rm disk}}\right)^{3}
\end{align} 

This will lead in turn to enhanced rates of migration in the disk, as discussed below.  

The models introduced in \citet{sirko2003}  do not prescribe an outer boundary of the disk. Such outer boundary is dictated by other constraints. In \citealt{sirko2003}, the consistency of the modeled spectral energy distribution with the observations determines the outer boundary. Given that no current mAGN has yet been identified, either because they exist for short timescales, exist only during early times of GCs formation or their properties are beyond the current observation capabilities we can not rely on observations, in this regard, and instead truncate the disk at $10^3 \ \rm{AU}$, which is a fraction of the average interparticle distance in the core of a GC (${\sim}1.3\times10^4\,\rm {AU}$ for a core number density of $5\times10^3 \,\rm pc^{-3}$). 
The mass of objects embedded in the disk could not exceed the mass of the disk, on the very extreme side, to maintain stability, which sets an upper bound on the number of objects in the disk. A more conservative and realistic assumption would be no more than $\sim 10\%$ of the disk mass. The mass of the disk up to the $10^3 \ \rm{AU}$ is approximately $100, 800$ and $4600 \ M_\odot$ for $M_{\rm IMBH} = 10^3, 10^4$ and $10^5 \, M_\odot$, respectively. Hence the disk can not host more than $\sim 10, 80$ and $460$ solar mass stars correspondingly or $\sim 1, 8$ and $46$ BHs with mass $10 \ M_\odot$.

\section{Motion in gas}\label{sec:motion in gas}

The dynamics of objects embedded in gas-rich environments deviate significantly from their evolution in a vacuum.
Several approaches can be employed to model this behavior, including mini-disks \citep{Stone2017}, migration within disks \citep{McKernan2012}, and modeling based on Bondi-Hoyle-Lyttelton (BHL) supersonic flows and their associated energy dissipation \citep{Antoni2019}.

The evolution within AGN disks is a complex and not fully understood phenomenon, with even the direction of migration torques remaining a subject of debate \citep{Moody2019,Duffel2020,Munoz2020,Grishin2023}. In this study, we will apply the conditions of an mAGN disk to investigate the evolution of objects within the disk under the influence of both type I/II migration and gas dynamical friction (GDF), comparing the results obtained from these two distinct approaches. Additionally, we will employ various density models, including a broken power-law model \citep{Grishin2023} based on the numerical solution of \citet{Gangardt2024}, as well as flat density profiles with varying densities.

\subsection{Gas-induced migration in gaseous disks}\label{subsec:disk-migration}

The migration of objects embedded within an accretion disk can also be modeled by utilizing the migration torques originally derived in the context of planet formation and evolution.
In this framework, migration is typically categorized into two regimes: type I and type II.
Type I migration is relevant for low-mass planets, where their interaction with the disk is relatively weak. In this regime, the torque driving migration arises from the gravitational interaction between the planet and the density waves it excites within the disk \citep{Ward1997,Tanaka2002,Paardekooper2010}.

The torque is given by 

\begin{align}
\Gamma_{\rm{I}} \approx - \left(\frac{m}{M_{\rm cen}}\right)^2\left(\frac{h}{r}\right)^{-2}\Sigma_g r^4 \Omega_K^2
\end{align}

\noindent
where $m$ is the mass of the migrating object, $M_{\rm cen}$ is the mass of the central object, $\Sigma_g$ is the surface density of the disk, $r$ is the radial distance from the central object, $\Omega_K$ is the Keplerian angular velocity and $h/r$ is the aspect ratio of the disk. Note that there might be some order of unity corrections due to the disk structure, which we will neglect. 

More massive planets could potentially open a gap in the disk and, for them, the torque is dominated by the viscous evolution of the disk as it interacts with the gap \citep{LinPapaloizou1979a,LinPapaloizou1979b,
LinPapaloizou1986,Ward1997,IdaLin2004,Duffell2014,Kanagawa2018}. In this regime, the torque is given by 

\begin{align}
\Gamma_{\rm{II}}\approx -3\pi \alpha \left(\frac{h}{r}\right)^2\Sigma_g r^4 \Omega_K^2
\end{align}

\noindent
where $\alpha$ is the Shakura Sunyaev parameter of the disk \citep{ShakuraSunyaev1973}. 

The total torque exerted on a migrating planet depends on the regime, such that 

\begin{align}
\Gamma_{\rm tot} = \begin{cases}
\Gamma_{\rm I} , \ R_{\rm Hill}<H\\
\Gamma_{\rm II}, \ R_{\rm Hill}>H
\end{cases}
\end{align}
where $R_{\rm Hill}= r \left(M_\star/3M_{\rm IMBH}\right)^{1/3}$ is the Hill radius
and the migration rate is given by 

\begin{align}
\frac{dr}{dt} = \frac{2\Gamma_{\rm tot}}{m r\Omega_K}
\end{align}

\begin{figure}%[H]
    \centering
    \includegraphics[width=1.3\linewidth]{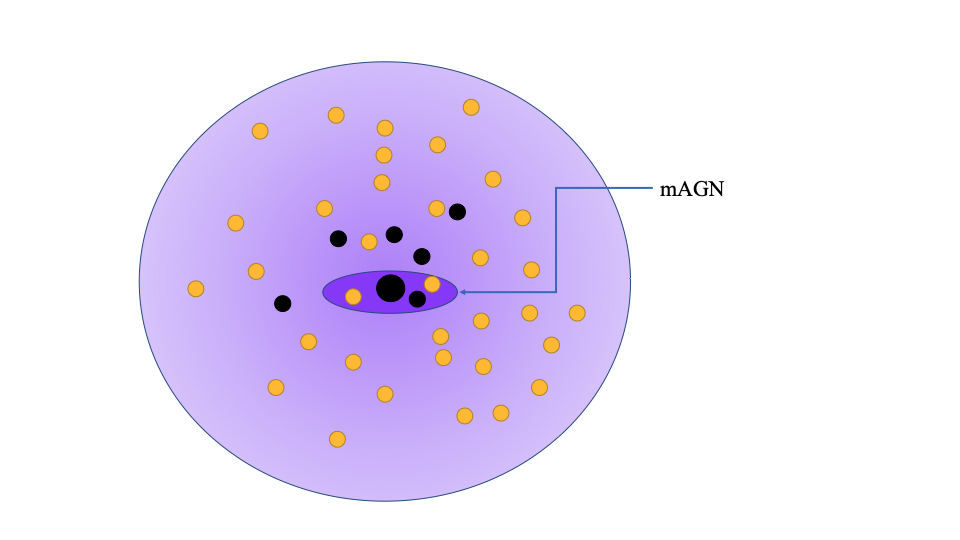}
    \caption{An illustration not to scale of the gas-rich cluster (in purple), the mini-AGN embedded in it (in dark purple), stars (in yellow) and black holes (in black).}
    \label{fig:illustration}
\end{figure}

\subsection{Gas dynamical friction}\label{sec:gas_dynamical_friction}
Here we use the gas dynamical friction (GDF) model \citep{Ostriker1999}. The interaction of an object with a nonzero velocity relative to the background gas dissipates energy from the binary and hence leads to hardening. The GDF force is given by 

\begin{equation}\label{eq:FGDF}
\textbf{F}_{\rm GDF}=-\frac{4\pi G^2 m^2 \rho_g}{v_{\rm rel}^3}\textbf{v}_{\rm rel}I\left(\frac{v_{\rm rel}}{c_s}\right)
\end{equation}

\noindent
where $m$ is the mass of the object, $\rho_g$ is the gas density, $v_{\rm rel}$ is the relative velocity between the gas and the object and $I$ is a function of $v_{\rm rel}$ and the speed of sound $c_s$, given by 

\begin{equation}
\begin{aligned}
I(\mathcal M) = 
\begin{cases}
\frac{1}{2}\ln \frac{1+\mathcal M}{1-\mathcal M}-\mathcal M, \ \mathcal M<1,\\
\frac{1}{2}\ln \frac{\mathcal M+1}{\mathcal M-1}+\ln \frac{\mathcal M-1}{r_{\rm min}/c_st}, \ \mathcal M>1
\end{cases}
\end{aligned}
\end{equation}

The separation evolution for a circular orbit is then given by \citep{MurrayDermott2000,GrishinPerets2016},

\begin{equation}
\frac{da}{dt}\bigg|_{\rm GDF}= \frac{2a^{3/2}}{\sqrt{G(m_1+m_2)}}\Delta f_{\rm GDF,\theta}
\end{equation}

\noindent
where $\Delta \textbf f_{\rm GDF}$ is the differential force per mass between the binary components, and $f_{\rm GDF, \theta}$ is its tangential component. 

\begin{equation}
\begin{aligned}
&\Delta \textbf{f}_{\rm GDF}\propto m_1\frac{\textbf{v}_{\rm rel,1}}{v_{\rm rel,1}^3}-m_2\frac{\textbf{v}_{\rm rel,2}}{v_{\rm rel,2}^3}, \\
&\textbf v_{\rm rel,1}= \textbf{V}_{\rm cm}+\frac{m_2}{m_1+m_2}\textbf{v}, \ 
\textbf v_{\rm rel,2}= \textbf{V}_{\rm cm}-\frac{m_1}{m_1+m_2}\textbf{v}
\end{aligned}
\end{equation}

\noindent
where $\textbf{V}_{\rm cm}$ is the center of mass velocity and $\textbf{v}$ is the relative velocity between the binary components $m_1$ and $m_2$. We consider the center of mass velocity $V_{\rm cm}$ to be the velocity dispersion of the binary and $v=v_{\rm Kep}$. In the supersonic regime, in which we focus, $\sigma\ll v_{\rm Kep}$ and hence one can neglect the contribution from $V_{\rm cm}$. Under these assumptions, the differential equation for the separation evolution could be reduced to 

\begin{equation}
\begin{aligned}\label{eq:a_evolution}
&\frac{da}{dt}\bigg|_{\rm GDF}= - \frac{8\pi G^{3/2}a^{3/2}}{\sqrt{m_1+m_2}}\rho_g(t)\frac{m_1}{v_{\rm Kep}^2}I \left(\frac{v_{\rm Kep}}{c_s}\right)\xi(q), \\ 
&\xi(q)=\left[(1+q^{-1})^2+q(1+q)^2\right] 
\end{aligned}
\end{equation}

\noindent
where $q=m_2/m_1$ is the mass ratio of the binary.
Note that this equation differs from eq. 7 in \cite{Rozner2023} by a factor of unity, due to mass-ratio corrections. 

This equation could be solved analytically, assuming that $I$ could be approximated by a constant for the motion, leading to the solution 

\begin{align}
&a(t)\approx \left(\frac{3}{2}\left[A\tau_{\rm gas}e^{-t/\tau_{\rm gas}}+C\right]\right)^{-2/3}, \\
\nonumber
&A = \frac{8\pi G^{1/2}}{(m_1+m_2)^{3/2}}\rho_{g,0}m_1I\xi(q), \\
\nonumber
& C = \frac{2}{3}a_0^{-3/2}-A\tau_{\rm gas} 
\end{align}

For very long timescales ($t/\tau_{\rm {gas}}\gg 1$), the solution converges towards an asymptote, 

\begin{align}
a(t\to \infty) = \left(\frac{3}{2}C\right)^{-2/3}
\end{align}

Moreover, it can be seen that the $\dot a|_{\rm GDF}\propto a^{2.5}$, such that once small enough separations are reached, GDF will become inefficient.

The typical timescale for a significant change due to GDF is given by 

\begin{align}
\tau_{\rm GDF} = \bigg|\frac{a}{\dot a|_{\rm GDF}}\bigg|\approx (Aa^{3/2})^{-1}
\end{align}

\noindent
where we assumed that this timescale is much shorter than the gas lifetime (which is justified later by the full solutions). 

\section{Dynamical processes}\label{sec:dynamical_processes}

Here we will sketch the dynamical processes induced by the presence of mAGN in a GC. We specify our fiducial parameters, in Table \ref{tab:fiducial values}, which are used throughout, unless stated otherwise.

\begin{table}
    \centering
    \begin{tabular}{|c|c|c|}
    \hline
        notation & name & fiducial value \\
        \hline
           $M_{\rm IMBH}$& IMBH mass & $10^4 \ M_\odot$\\
          $R_{\rm mAGN}$ & radius of the mAGN & $1-2\times 10^3 \ \rm{AU}$\\
                   $M_{mAGN}$ & mAGN mass & $1200 \ M_\odot$ \\
         $h/r$ & aspect ratio & $0.01-0.05$\\
         $\alpha$ & viscosity parameter & $0.01$\\
         $\tau_{\rm gas}$ & gas lifetime & $50 \ \rm{Myr}$\\
                  \hline
    \end{tabular}
    \caption{The fiducial values used in the paper unless stated otherwise. }
    \label{tab:fiducial values}
\end{table}

\subsection{Gas hardening}\label{subsec:gas-hardening}

Both disk migration and GDF extract energy from the binary, leading to a separation decrease, i.e. gas hardening of the binary. 

\begin{figure*}
    \centering
    \includegraphics[width=1.\linewidth]{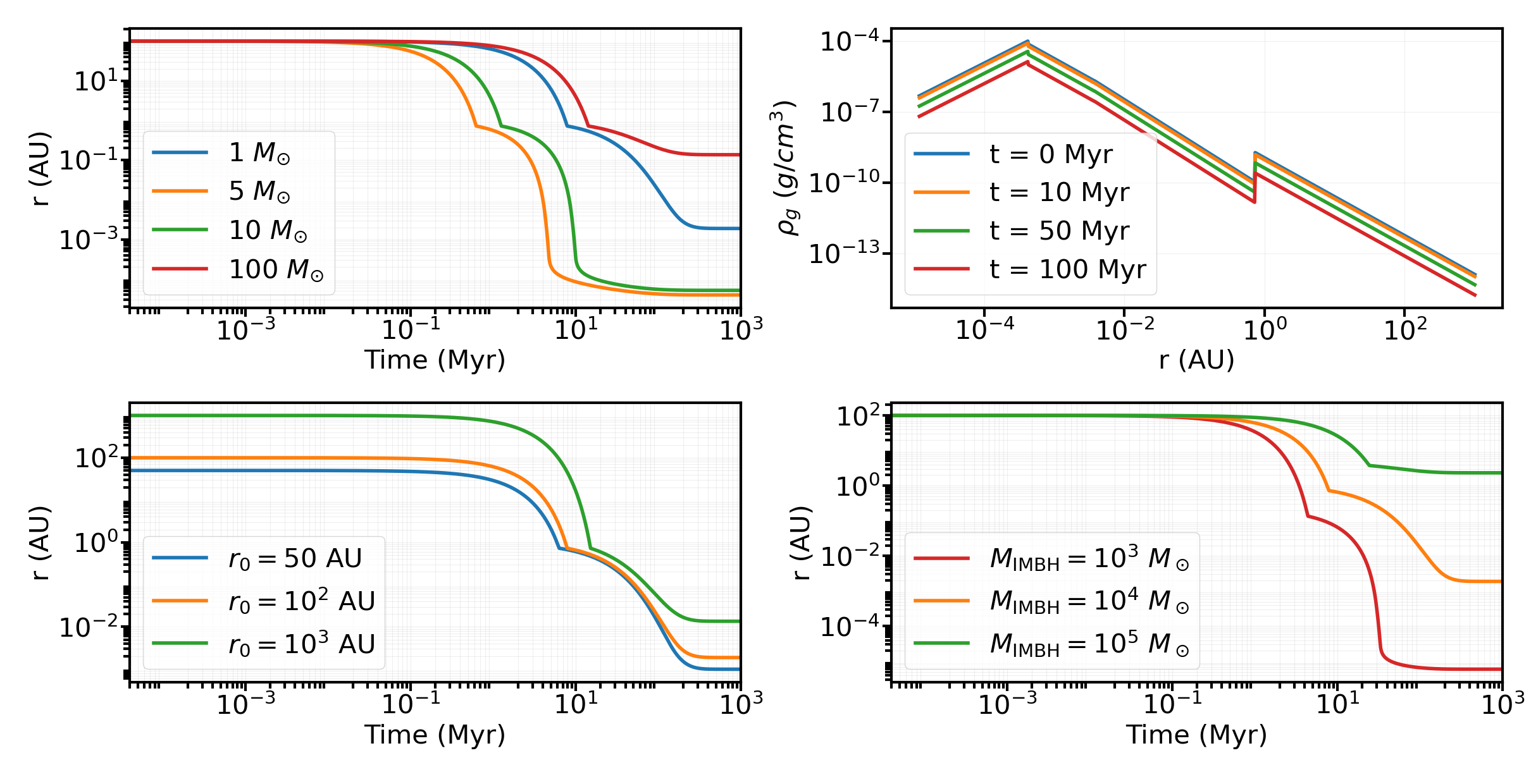}
    \caption{The evolution of objects under the effect of \textit{disk migration} only, in a mAGN disk with a radius of $1000 \ \rm{AU}$, $\alpha=0.01$, $h/r=0.05$ and typical gas lifetime of $50 \ \rm{Myr}$ with an exponential decay. Unless stated otherwise, the default parameters that we vary are $M_{\rm IMBH}$, gas density profile as described in \ref{fig:Sirko}, initial separation of $10^2 \ \rm{AU}$ and solar mass migrating star. Upper left: Evolution in time of the separation between the IMBHs and objects of different masses. Upper right: Evolution in time of the gas density profile described in Fig. \ref{fig:Sirko} for an IMBH of mass $10^4 \ M_\odot$. Lower left: Evolution in time of the separation, for different given separations (note the different y-axis). Lower right: Migration for different masses of the IMBH.}
    \label{fig:disk-migration}
\end{figure*}

Figure \ref{fig:disk-migration} illustrates the migration of objects within an mAGN disk under the influence of disk migration. The upper left panel depicts the time evolution of the separation between the IMBH and objects of varying masses. It is evident that the evolutionary tracks diverge significantly for different masses, with more massive objects exhibiting faster migration rates. Furthermore, after a while, the separation reaches an equilibrium value that is mass-dependent. The upper right panel showcases the time evolution of the density profile, assuming an exponential decay with a characteristic timescale of $50 \ \rm{Myr}$. The lower left panel presents the separation evolution for various initial separations, revealing a weak dependence on the initial conditions. The lower right panel demonstrates the evolution of a solar-mass star within mAGN disks surrounding IMBHs of different masses. Notably, higher IMBH masses result in faster migration rates and distinct final separations. Across all the evolutionary plots, multiple evolution regimes are observed, governed by migration into different density regions of the disk (refer to the upper right panel for the density profile). The rapid migration of objects with diverse masses towards a common small separation could potentially catalyze collisions and trigger the formation of transient phenomena.

\begin{figure*}
    \centering
    \includegraphics[width=1.\linewidth]{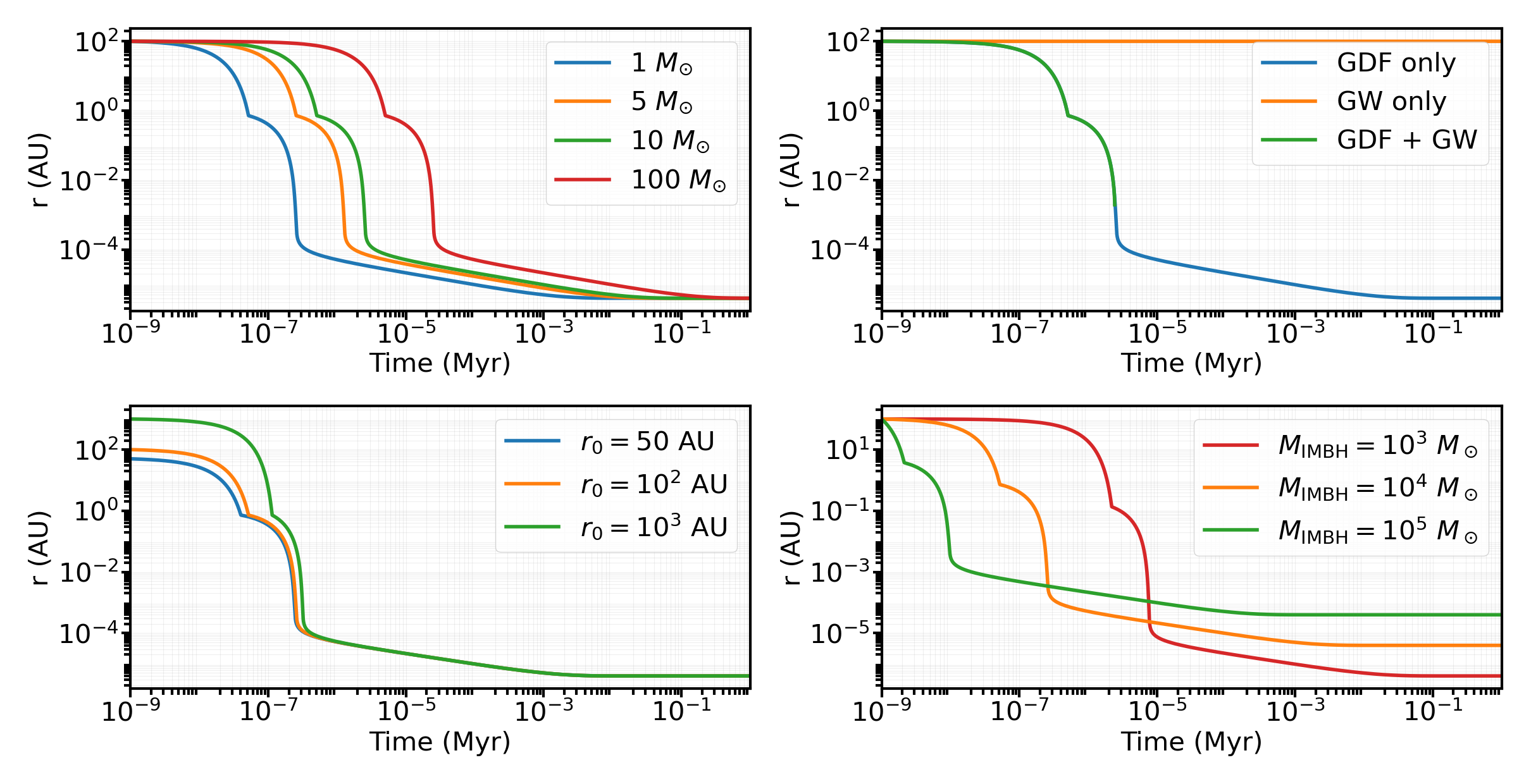}
    \caption{The evolution of objects under the effect of \textit{gas dynamical friction} only, in a mAGN disk with a radius of $1000 \ \rm{AU}$, $\alpha=0.01$, $h/r=0.05$ and typical gas lifetime of $50 \ \rm{Myr}$ with an exponential decay. Unless stated otherwise, the default parameters that we vary are $M_{\rm IMBH}$, gas density profile as described in \ref{fig:Sirko}, initial separation of $10^2 \ \rm{AU}$ and solar mass migrating star. Upper left: Evolution in time of the separation between the IMBHs and objects of different masses. 
    Upper right: comparison between the evolution under GDF only, GW only and the combined effect of the two, for a $10 \ M_\odot$ object migrating in a disk surrounding a $10^4 \ M_\odot$ IMBH. 
    Lower left: Evolution in time of the separation, for different given separations. Lower right: Migration for different masses of the IMBH.}
    \label{fig:GDF}
\end{figure*}

Figure \ref{fig:GDF} illustrates the evolution of objects within the mAGN disk solely under the influence of gas dynamical friction (GDF). The upper left panel displays the time evolution of separation for objects of varying masses. It is apparent that GDF is a highly efficient mechanism, causing a substantial separation shrinkage over remarkably short timescales, potentially even less than a year. Interestingly, more massive objects exhibit slower migration, although this migration saturates after a brief period as the separation becomes very small. This diminished separation results in a high relative velocity between the object and the gas, rendering GDF less effective (since $F_{\rm GDF}\propto v_{\rm rel}^{-3}$).

The upper right panel underscores the crucial role of GDF in gravitational wave (GW) emission. In the absence of gas, no GW emission is anticipated under these conditions. However, the introduction of gas leads to an expected merger within a relatively short timeframe. The lower left panel presents the evolutionary tracks for objects starting from different initial separations. Due to the high efficiency of GDF, the dependence on the initial separation is weak, with all curves converging after a short duration. The lower right panel showcases the evolution of a star within disks surrounding IMBHs of varying masses. It is observed that higher IMBH masses correspond to more efficient migration.

\subsection{Alignment into the disk}\label{subsec:alignment}

Consider a star with mass $m_\star$ in an inclined orbit with respect to the disk. The typical timescale to dissipate the velocity with gas dynamical friction could be estimated by (e.g. \citealp{Artymowicz1993,
Rein2012,Bartos2017,Panamarev2018,Fabj2020,
GenerozovPerets2022,Nasim2022}),

\begin{equation}
\tau_{\rm align}\sim
\frac{v_{\rm Kep}^3(a)}{G^2 \rho_g m \frac{h}{r}\ln \Lambda}\sin^3 \frac{i}{2}\sin i
\end{equation}

\begin{figure}%[H]
    \centering
    \includegraphics[width=1.1\linewidth]{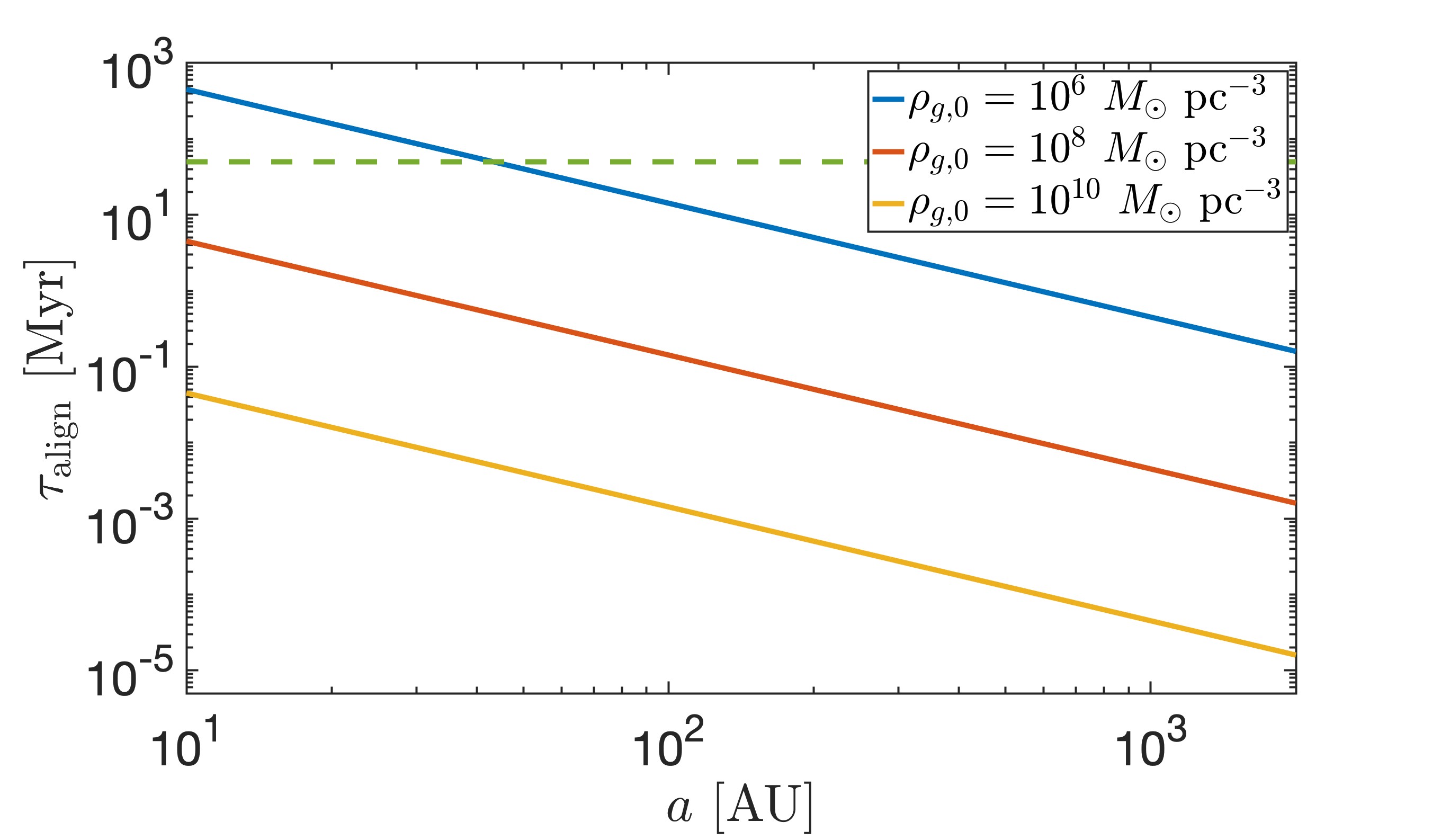}
    \caption{Typical alignment timescales for a solar mass star with an inclination of $\pi/2$, for different gas densities (solid lines), in comparison to the gas lifetime (dashed line). These densities correspond to $7\times 10^{-17}, \ 7\times 10^{-15}$ and $7\times 10^{13} \ \rm{g/cm^3}$ correspondingly. }
    \label{fig:alignment_time_density_1msun}
\end{figure}

\noindent
Figure \ref{fig:alignment_time_density_1msun} presents the alignment timescales for solar-mass objects as a function of gas density and semi-major axis. For typical gas densities, the alignment timescales with the disk are short, often shorter than the gas lifetime itself. This suggests that all or the vast majority of solar-mass objects are expected to align with the disk plane.

However, the number of objects that can align with the disk is limited by stability considerations related to the disk's mass, as discussed earlier. In general, larger disks capable of hosting a sufficient number of stars may allow massive perturbers to align via resonant dynamical friction \citep{GinatKocsis2023}.

\subsection{Gravitational wave mergers}\label{subsec:GW_rate}

Gas hardening, the process by which the energy of binaries dissipates more rapidly in the presence of gas, can act as a catalyst for gravitational wave (GW) mergers. This phenomenon has been demonstrated in the context of gas-rich globular clusters \citep{RoznerPerets2022} and other gaseous environments like AGN disks \citep{McKernan2012,Stone2017,Tagawa2020}.

Mini-AGNs, by their nature, could harbor binaries comprising the IMBH and a companion object. These binaries have the potential to become sources of intermediate-mass ratio inspirals (IMRIs) or, in the case of massive IMBHs, extreme-mass ratio inspirals (EMRIs). Traditionally, EMRIs are thought to arise from two-body encounters (two-body relaxation) at the centers of dense stellar nuclei, tidal-induced binary separation, or migration \citep{HopmanAlexander2005,Amaro-Seoane2018,Miller2005,RavehPerets2021,Levin2007}. However, within gas-rich media, alternative formation channels become viable.

There are several unique features to the GWs channel described above. 
The effect of gas hardening could shift the frequency of the produced GWs (see a discussion on frequency shifting in a different context in \citealp{Tanaka_2008,Hendriks2024,Samsing2024}),

\begin{align}
\frac{\Delta f}{f}\approx \frac{T_{\rm obs}}{\tau_{\rm merge}}
\end{align}

\noindent
where $f$ is the observational frequency band, $T_{\rm obs}$ is the duration of the observation, which is typically one year, and $\tau_{\rm merge}\approx 20 \ \rm{yr}$ is the typical merger timescale for $\rho_{g,0} = 10^{10} \ M_\odot \ \rm{pc}^{-3}$

The frequency shift translates to a shift in the number of cycles before the plunge,

\begin{align}
\Delta N \approx \Delta f T_{\rm obs} = fT_{\rm obs}\frac{T_{\rm obs}}{\tau_{\rm merge}}\approx 1577
\end{align}

\noindent
for $f=10^{-3} \ \rm{Hrz}$, which is the observational frequency of LISA.  
Since the typical number of cycles is $N\sim f T_{\rm obs}\sim 31557$, the number of extra orbits is significant ($\sim 5\%$ increment). For smaller gas densities, such as $\rho_{g,0}=10^6 \ M_\odot \ \rm{pc}^{-3}$, $\tau_{\rm merge}\approx 2\times 10^{-2} \ \rm{Myr}$ which translates to a shift of $\Delta N \approx 1.6$, which in principle could be observed as well. 
A larger number of cycles could improve our observational understanding of the shape of these GWs.

IMBH-BH mergers could potentially be highly misaligned unless there is a spin-up induced by accretion that leads to alignment. Misaligned mergers could lead to high-velocity GW kicks. Strong enough kicks could eventually deplete the GC from IMBHs.

The existence of mAGN disks could also increase the production rates of GW sources, and specifically the rates of IMRIs, and hereby we will estimate the BH-IMBH merger rate, considering a BH with mass $10 \ M_\odot$ and an IMBH with mass $10^4 \ M_\odot$. 
This rate is sensitive to the assumptions on several distributions, as we will describe below.
The rate is then given by

\begin{equation}
\Gamma\sim f^{\rm GC}_{\rm IMBH}\frac{n_{\rm GC}N_{\star}f_{\rm mAGN}f_{\geq 20 M_\odot}f_{\rm ret}f_{\rm merge}}{\tau_{\rm GC}}
\end{equation}

\noindent 
where $f^{\rm GC}_{\rm IMBH}$ is the fraction of GCs that host an IMBH, $n_{\rm GC}$ is the number density of GCs,
$N_\star$ is the number of stars in the cluster, $f_{\rm mAGN}$ is the fraction of stars from the cluster that reside in the mAGN disk, $f_{\geq 20 M_\odot}$ is the fraction of stars with masses larger than $20 \ M_\odot$, $f_{\rm ret}$ is the retention fraction in the cluster and $f_{\rm merge}$ is the fraction of binaries that merge. We take the lifetime of the GC as $10 \ \rm{Gyr}$.

The fraction of GCs harboring an IMBH is highly uncertain, contingent upon various factors including IMBH formation and retention within the cluster. Following \citet{Fragione2022,Tang2024}, we adopt an occupation fraction ($f^{\rm GC}_{\rm IMBH}$) ranging between $0.01-0.1$. 
The number density of GCs is taken to be within the range of $(0.33 - 3) \times 10^9 \ \rm{Gpc}^{-3}$, in accordance with \citet{KritosCholis2020}. 

Stability considerations dictate that the total number of black holes within the mAGN cannot surpass $10$, implying $N_{\star} f_{\rm mAGN}f_{\gtrsim 20 M_\odot}f_{\rm ret}\leq 10$. For the lower bound, we factor in the alignment of binaries into the disk from the surrounding sphere. As illustrated in Figure \ref{fig:alignment_time_density_1msun}, the alignment timescales are sufficiently short to consider the entire sphere around the mAGN disk, as objects will eventually align within the disk's lifetime. This translates to $f_{mAGN}\sim \left(R_{\rm mAGN}/R_{\rm GC}\right)^3\sim 10^{-7}$ for $R_{\rm mAGN}\sim 10^3 \ \rm{AU}$ and $R_{\rm GC}\sim \rm{pc}$.

We employ a Kroupa mass function \citep{Kroupa2002} for the cluster, resulting in a fraction of stars with masses exceeding $20 \ M_\odot$ of $2\times 10^{-3}$ in a non-segregated environment. The retention fraction within the cluster is assumed to be $10\%$ \citep{KritosCholis2020}. Based on our model, we set the merger fraction ($f_{\rm merge}$) to $1$. Under these assumptions, the merger rate resulting from our proposed channel ranges between $(\sim 7 \times 10^{-10}-0.03) \ \rm{Gpc}^{-3} \ \rm{yr}^{-1}$. It should be noted that even under optimistic assumptions, one can restrict the number of mergers to not exceed one merger during the lifetime of the mAGN, due to GW kicks that can destroy the disk.
Hopefully, with future measurements, some of the uncertainties will be cleared, leading to a more constraining rate calculation. This range can potentially contribute to the typically expected observational rates of $0.1-10 \ \rm{Gpc}^{-3} \ \rm{yr}^{-1}$ \citep{Abbott2017,Abbott2019,Abbott2022,Fragione2022}.

\subsection{Tidal disruption events (TDEs)}\label{subsec:TDE}

The dynamics in the mAGN could lead to the disruption of stars by the IMBH, producing a unique type of TDEs, with typical masses $10^{-3}-10^{-2}$ smaller than the masses of the 'standard' TDEs produced by interactions with MBHs, and hence we term as mTDEs and cTDEs correspondingly. The tidal disruption radius is given by 

\begin{equation}
R_t = R_\star \left(\frac{M_{\rm IMBH}}{M_\star}\right)^{1/3} 
\end{equation}

\noindent
which is $22 \ R_\odot$ approximately for $R_\star=R_\odot, \ M_\star=M_\odot$ and $M_{\rm IMBH}= 10^4 \ M_\odot$. For a mTDE to occur, the disrupted object should pass by the IMBH at a distance that does not exceed the tidal disruption radius. Such a close encounter could take place either by a random close encounter or a gas-assisted encounter.  

The most bound material first returns to the pericenter after a typical timescale given by (e.g. \citealp{Rees1988,Perets2016muTDE})

\begin{equation}
\begin{aligned}
&t_{\rm min}=\frac{2\pi R_t^3}{(GM_{\rm IMBH})^{1/2}(2R_\star)^{3/2}}\approx 
\\ \nonumber
&\approx 3\times 10^5 \ \rm{sec} \left(\frac{R_\star}{R_\odot}\right)^{3/2}\left(\frac{M_{\rm{IMBH}}}{10^4 \ M_\odot}\right)^{1/2}\left(\frac{M_\odot}{M_\star}\right)
\end{aligned}
\end{equation}

\noindent
The bound material then returns to the pericenter at a rate given by 

\begin{equation}
    \dot M(t)=\frac{1}{3}\frac{M_\star}{t_{\rm min}}\left(\frac{t}{t_{\rm min}}\right)^{-5/3}
\end{equation}

\begin{figure}%[H]
    \centering
    \includegraphics[width=1\linewidth]{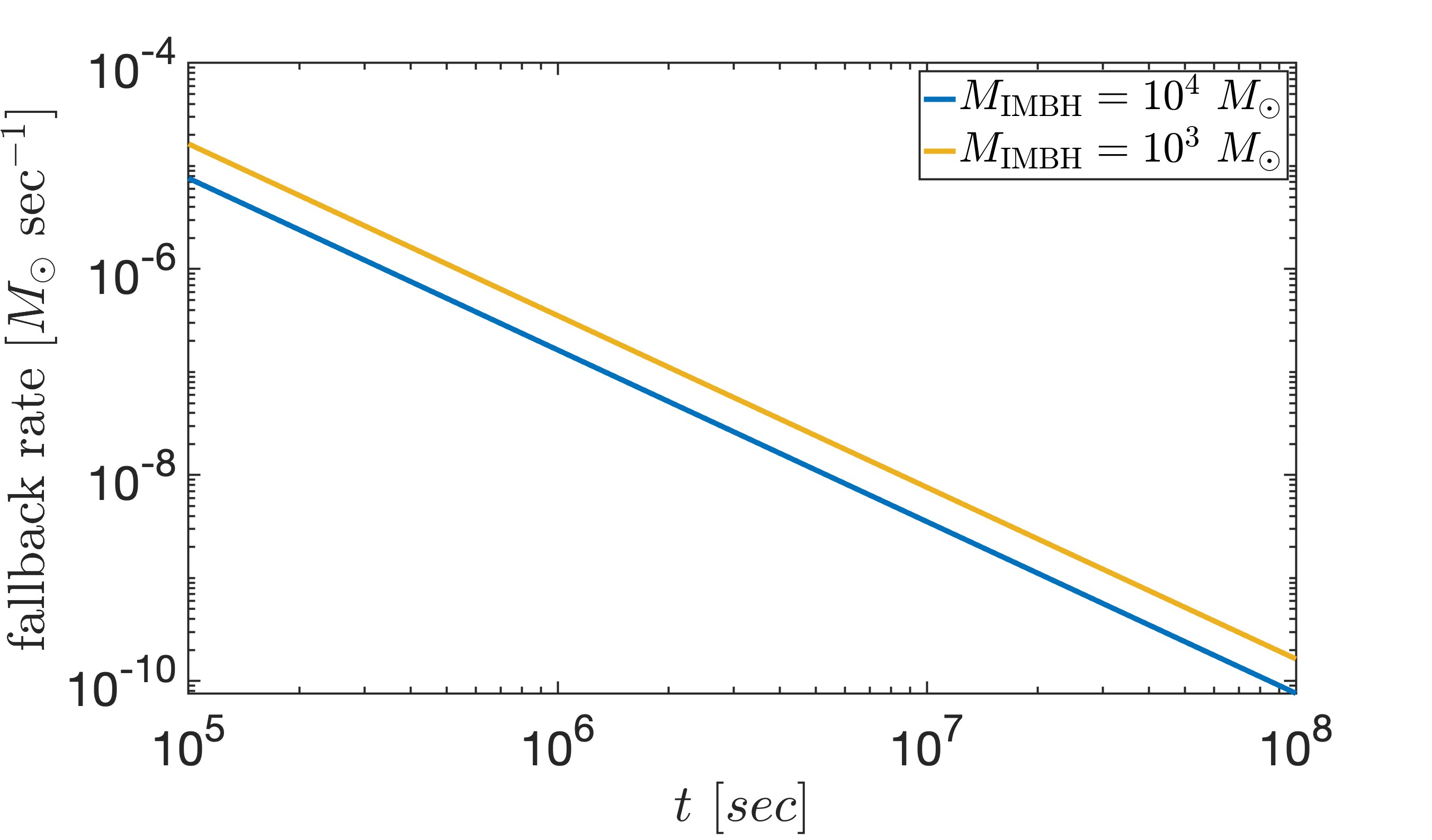}
    \caption{Approximate late times rate for a solar mass star and an IMBH with masses of $10^3/10^4 \ M_\odot$. }
    \label{fig:enter-label}
\end{figure}

\noindent
Generally, $dM/dE$, i.e. mass per unit energy, varies with the strength of the tidal interaction and the density
profile of the disrupted objects. For completely disrupted objects, this dependence is relatively flat, such that for late times, $
\dot M\propto t^{-5/3}$. 

The event rate of cTDEs per cluster could be estimated by 

\begin{align}
\Gamma_{\rm{cTDE}}\approx N_{\rm {IMBH}} \frac{2 \pi G (M_{\rm{IMBH}}+M_\star)N_\star R_t }{\sigma V_c} \approx  0.18 \ \rm{Myr}^{-1}
\end{align}

\noindent
where $N_{\rm{IMBH}}$ is the number of IMBHs in the cluster, $N_\star$ is the number of stars, $R_t= R_\star\left(M_{\rm {IMBH}}/M_\star\right)^{1/3}$ is the tidal radius and $V_c\approx 4\pi R_{\rm{cl}}^3/3$ is the volume of the cluster. Here we assume that the cross-section is dominated by gravitational focusing and that in practice the TDE rate is determined by the collision rates of IMBHs. Here we considered $M_{\rm{IMBH}}= 10^4 \ M_\odot, \ M_\star = M_\odot, \ N_\star = 10^5, \ R_\star = R_\odot, \ \sigma = 20 \ \rm{km/sec}, R_{\rm{cl}}=\rm{pc}$.   

The peak luminosity of a cTDE is expected to be the Eddington luminosity, and hence the typical peak luminosity of a cTDE is expected to be $10^{-3}-10^{-2}$ weaker than the one of a standard TDE. It should be noted that while cTDE could be just limited by Eddington, in some cases, we might get a jet that might be far more luminous (see discussion in the context of $\mu$TDE in \citealp{Perets2016muTDE}). 

\section{Formation of Accreting IMBHs from Captured  Binaries}\label{sec:accreting_IMBH}

The presence of an IMBH embedded in a gas-rich environment,
as envisioned in our proposed mAGN scenario, presents the opportunity for the formation of accreting systems akin to ultraluminous X-ray sources (ULXs). This process can be initiated when stars or binaries migrate through the disk and are subsequently captured by the IMBH, potentially resulting in mass transfer and accretion. Currently, several lines of evidence suggest the presence of IMBHs in ULXs (for a comprehensive review, refer to \citet{Kaaret2017} and references therein).

Below we discuss the formation of accreting IMBHs from binary companions captured due to migrating stars in the AGN disk.

\subsection{Formation of Accreting Systems}

Once a binary system is captured by the IMBH, it can undergo evolution into an accreting system via multiple pathways:

1. \textbf{Direct capture of a binary:} In the event of a tight binary being captured, the less massive component may expand to fill its Roche lobe, leading to mass transfer onto the IMBH and the establishment of an accreting system \citep{Hopman2004}.

2. \textbf{Tidal capture of a single star:} A single star can be tidally captured by the IMBH. Subsequently, the star may evolve to fill its Roche lobe, initiating mass transfer and accretion onto the IMBH \citep{Hopman2004, Patruno2006}.

The typical rate for such a tidal capture per an IMBH could be estimated by 

\begin{align}
&\Gamma_{tc}=n \pi R_t^2 \left(1+\frac{2G M_{\rm{IMBH}}}{R_t \sigma_\star^2}\right) \sigma_\star\approx 10^{-5} \ \rm{Myr}^{-1}, \\
&R_t = R_\star \left(\frac{M_{\rm {IMBH}}}{m_\star}\right)^{1/3}
\end{align}

\noindent
when we considered the parameters of a star with a solar mass and a solar radius, $\sigma_\star=10 \ \rm{km/sec}$ and $M_{\rm IMBH}=10^4 \ M_\odot$. It should be noted that the cross-section can change due to the presence of gas. 

3. \textbf {Exchange interactions:} In a dense environment, the IMBH may acquire a stellar companion through exchange interactions with existing binaries \citep{Mapelli2013}.

\subsection{Accretion and ULX Formation}

The accretion rate onto the IMBH in these captured systems has the potential to reach or even surpass the Eddington limit, resulting in the formation of an object resembling an ultraluminous X-ray source (ULX) \citep{Hopman2004}. The accretion rate is primarily determined by the mass transfer rate from the companion star, which is influenced by factors such as the stellar type, orbital parameters, and the evolutionary stage of the donor star.

\citet{Hopman2004} demonstrated that the mass transfer rates in such systems can be sustained at super-Eddington levels for extended periods, potentially offering an explanation for the observed luminosities of ULXs. Subsequent studies have built upon this work, incorporating the effects of disk winds \citep{Poutanen2007} and investigating the parameter space that allows for stable mass transfer \citep{Frago2015}.

\subsection{Observational Implications}
The formation of accreting IMBHs within mAGN disks could give rise to several observable consequences:

1. \textbf{X-ray emission:} These systems are expected to generate X-ray emission, potentially detectable as point sources within globular clusters or contributing to the overall X-ray luminosity of the cluster \citep{Maccarone2007}.

2. \textbf{Optical signatures:} The accretion disk and the irradiated companion star could produce detectable optical emission, particularly in nearby clusters \citep{Patruno2006}.

3. \textbf{Transient phenomena:} Fluctuations in the accretion rate could lead to transient behavior, including state transitions and outbursts \citep{Kaaret2017}.

The detection of such accreting IMBHs within globular clusters would provide compelling evidence for the existence of IMBHs and lend support to the mAGN scenario proposed in this study. However, differentiating these sources from other types of X-ray binaries may necessitate high-resolution X-ray observations and multi-wavelength follow-up investigations.

\section{Large-scale Jets from Gas-embedded IMBHs in GCs}\label{sec:jets}

The presence of gas-embedded IMBHs in GCs raises the intriguing possibility of observable large-scale jets. While jets are commonly associated with AGNs powered by supermassive BHs, recent studies suggest that IMBHs might also be capable of launching significant outflows under certain conditions.

\subsection{Jet Formation Mechanism}

The formation of jets from accreting BHs is generally attributed to the interaction between the accretion disk's magnetic field and the spinning black hole, as described by the Blandford-Znajek mechanism \citep{Blandford1977}. For IMBHs in GCs, the key requirements for jet formation would be sufficient gas supply for accretion, the presence of a magnetic field in the accretion disk, and rotation of the IMBH. 

\citet{Maccarone2004} proposed that IMBHs in GCs could indeed produce radio jets, albeit at lower luminosities compared to AGN jets. The expected radio luminosity would scale with the BH mass and accretion rate, following the fundamental plane of BH activity \citep{Merloni2003}.

The typical power of such a jet could be estimated by 

\begin{align}
P_{\rm jet}\approx \epsilon \dot M_{\rm acc}c^2\approx 10^{42} \ \frac{\rm erg}{\rm sec}
\end{align}

\noindent
where $\epsilon$ is the efficiency factor of accretion taken to be $0.1$, $\dot M_{\rm acc}$ is the rate of mass accretion on the IMBH and $c$ is the speed of light. Accretion at the Eddington rate could be estimated by 

\begin{align}
\dot M_{\rm acc}=\frac{L_{\rm Edd}}{\eta c^2}
\end{align}

\noindent
where $\eta$ is an efficiency factor, taken to be $0.1$. 

The Eddington luminosity is given by 

\begin{align}
L_{\rm Edd}= \frac{4\pi GM_{\rm IMBH}m_pc}{\sigma_T}\approx 10^{42}\frac{\rm{erg}}{\rm{sec}}\left(\frac{M_{\rm IMBH}}{10^4 \ M_\odot}\right) 
\end{align}

\noindent
where $m_p$ is the proton mass and $\sigma_T$ is the Thompson scattering cross-section for electrons.  

\subsection{Observational Prospects}

Detecting jets from IMBHs in GCs presents several challenges:

\begin{itemize}
    \item \textbf{Low luminosity:} The relatively low mass of IMBHs compared to SMBHs translates to less powerful jets. 
    \item \textbf{Intermittent activity:} The limited gas supply in GCs could lead to episodic accretion and jet formation, making detection more difficult. 
    \item \textbf{Resolution:} High angular resolution is necessary to resolve jets on the scale of GCs.
\end{itemize}

Despite these challenges, \citet{Mezcua2018} suggested that deep radio observations with current and future facilities, such as the Square Kilometre Array (SKA) and next-generation Very Large Array (ngVLA), could potentially enable the detection of jets from IMBHs in the mass range of $10^3 - 10^5 M_\odot$. 

\subsection{Implications and Future Prospects}

The detection of large-scale jets from gas-embedded IMBHs in GCs would have profound implications:

\begin{itemize}
    \item It would provide compelling evidence for the existence of IMBHs within GCs.
    \item It would confirm the presence of accretion disks around these IMBHs.
    \item It would offer valuable insights into the scaling of jet properties across a vast range of black hole masses.
    \item It could potentially contribute to explaining the observed multiple stellar populations in GCs, as jets might influence the cluster's gas dynamics and star formation history \citep{Krause2013}.
\end{itemize}

Future multi-wavelength observations, combining radio, X-ray, and possibly gamma-ray data, will be essential in the search for and characterization of jets from IMBHs in GCs. Such discoveries would open a new window into the physics of intermediate-mass black holes and their impact on the evolution of globular clusters.

\section{Discussion}\label{sec:discussion}

\subsection{Few-body dynamics}\label{subsec:few-body dynamics}

The presence of gas within an mAGN can potentially induce binary formation \citep{Tagawa2020,Rozner2023,Whitehead2023,DodiciTremaine2024}. This process occurs when an object penetrates the Hill sphere of another object and dissipates enough energy through GDF to become gravitationally bound, forming a new binary.

For a solar-mass star orbiting an IMBH of $10^4 M_\odot$ within a disk of $10^3 \ \rm{AU}$ radius, the typical Hill radius is approximately 30 AU. Assuming a stellar number density of $n_\star \sim 10^3 \ \rm{pc}^{-3}$ and a velocity dispersion of $\sigma\sim 10 \ \rm{km/sec}$, the capture rate into the disk is estimated to be $nR_{\rm Hill}^2 \sigma \sim 2\times 10^{-5} \ \rm{Myr}^{-1}$, indicating that such captures are relatively infrequent. 

Nevertheless, captures into the mAGN could potentially lead to interactions between stars within the disk, in addition to the previously discussed phenomena. These interactions might include binary-single encounters \citep{Hut1983,HutBahcall1983,Samsing2014,StoneLeigh2019,GinatPerets2021,Samsing2022,Trani2024} or dynamical events involving higher-order hierarchies. Furthermore, gas-assisted binary formation could also be triggered in principle \citep{Rozner2023}.

\subsection{Survival of mAGN disks}\label{subsec:lifetime}

The processes described earlier could be disrupted by the ejection of the IMBH or the destabilization of the disk.

Two primary channels could lead to IMBH ejection: recoil kicks from BH mergers and strong gravitational scatterings. Anisotropic gravitational wave emissions from BH mergers can impart high-speed kicks to IMBHs, potentially reaching velocities of $\sim 4\times 10^3 \ \rm{km/sec}$ \citep{Merritt2004} , and potentially breaking the disk after one merger. Given that the characteristic velocity dispersion in gas-free GCs is roughly 50 km/s, a significant fraction of IMBHs is expected to be ejected from the cluster, with the exact fraction depending on the mass ratio \citep{Atallah2023}.

The typical lifetime of the disk is influenced not only by the gas accretion rate but also by the capture of objects. The disk's stability is compromised once it captures a comparable mass of objects, establishing an upper limit on its lifetime.

\subsection{Brownian motion }\label{subsec:Brownian}

The motion of IMBHs in GCs could be described as a random walk in the momentum-space, as its motion is perturbed by gravitational encounters with other celestial objects, similar to other dense environments \citep{Merritt2006,DiCintio2023}. 

The velocity amplitude is given by

\begin{align}
\sigma_{\rm IMBH} \approx 0.07 \ \frac{\rm{km}}{\rm{sec}}\left(\frac{\sigma_\star}{10 \rm{km/sec}}\right)\left(\frac{m_\star}{0.5 \ M_\odot}\frac{10^4 \ M_\odot}{M_{\rm IMBH}}\right)^{1/2}
\end{align}

\noindent
where $\sigma_\star$ and $m_\star$ are the velocity dispersion and mass of the background stars and $M_{\rm IMBH}$ is the mass of the IMBH. 

The amplitude of the Brownian motion can be estimated by 

\begin{align}
\braket{x^2}\approx 2Dt\approx 2\frac{\sigma^2_{\rm IMBH}}{t_{\rm coll}}t\approx 10^{-3} \rm{AU} \frac{t}{1 \ \rm{Myr}}
\end{align}

\noindent
where $D$ is the diffusion coefficient of the motion, and $t_{\rm coll}$ is the typical collision time given by 

\begin{align}
t_{\rm coll} = \frac{1}{n\sigma_\star \pi R_\star^2}\approx 6340 \ \rm{yrs}
\end{align}

\noindent
when we considered $\sigma_\star = 10 \ \rm{km/sec}, \ R_\star = R_\odot$.
Hence, the overall displacement is not expected to significantly change the location of the IMBH and its accretion disk. 

\subsection{Observational Implications of mAGNs}
The existence of mAGNs within globular clusters (GCs) opens up a new avenue for observational exploration and offers unique opportunities to test our understanding of IMBHs and their interaction with their environment. Indeed, there might have already been some potential indications for such phenomena (e.g., \citealp{Filippenko2003}).

\subsubsection{Electromagnetic Signatures}
The accretion process onto the IMBH within the mAGN disk is expected to produce luminous electromagnetic emission across a wide range of wavelengths. X-ray observations could reveal the presence of the mAGN through its characteristic thermal emission and potential variability \citep{Maccarone2007}. Additionally, radio observations could unveil the presence of jets emanating from the mAGN, providing further evidence of accretion onto the IMBH \citep{Mezcua2018}. The detection of these electromagnetic counterparts would not only confirm the existence of mAGNs but also offer valuable insights into the accretion physics of IMBHs.

\subsubsection{Gravitational Wave Signals}
The dynamical processes occurring within mAGNs, such as the inspiral and merger of stellar-mass black holes and IMRIs with the central IMBH, are expected to generate gravitational waves (GWs) detectable by current and future GW observatories. The detection of IMRIs and EMRIs with unique mass ratios, as discussed in Section~\ref{subsec:GW_rate}, would offer a direct probe of IMBH demographics and their formation channels. The observation of GWs from the disruption of stars by the IMBH, or mTDEs, as outlined in Section~\ref{subsec:TDE}, could shed light on the tidal disruption process and the properties of the disrupted star.

\subsubsection{Unique Transients}
The capture and disruption of stars by the IMBH within the mAGN disk could give rise to various transient phenomena, such as mTDEs and cTDEs (see Section~\ref{subsec:TDE}). These events, characterized by sudden bursts of electromagnetic radiation, could provide additional evidence of the presence of IMBHs in GCs. The study of these transients would allow us to probe the properties of the disrupted star and the accretion physics of the IMBH.

\subsubsection{Variability}
The interaction of stars and binaries with the mAGN disk could lead to variability in the disk's structure and accretion rate, resulting in observable variations in the electromagnetic emission from the mAGN. Monitoring the variability of mAGNs could provide valuable information about the dynamics of stars within the disk and the accretion process onto the IMBH, complementing the studies of gas hardening presented in Subsection \ref{subsec:gas-hardening}.

\subsubsection{Additional Observational Aspects}
As mAGNs are expected to form mostly during the early gas-rich phases of GCs, it would be interesting to search for them in young massive clusters, which are believed to be the progenitors of GCs. Detecting mAGNs in these environments would provide crucial insights into the early evolution of IMBHs and their host clusters. Combining observations across different wavelengths, such as X-ray, radio, and optical, would offer a comprehensive view of mAGNs and their diverse observational signatures. This would enable a more detailed characterization of the accretion process, the dynamics of stars within the disk, and the potential feedback effects on the host cluster, as hinted at in Subsection \ref{subsec:few-body dynamics}.

\section{Caveats and future directions}\label{sec:caveats}

Here we briefly mention some of the possible caveats of our work:

\begin{itemize}
    \item While the existence of second-generation stars in GCs is commonly observed, their formation scenario and the associated gas lifetime, amounts, and distributions remain highly uncertain. In this study, we operated under the assumption that there was at least the amount of gas required to form the second generation. However, alternative channels have been proposed to explain their existence. It's worth noting that our model could be inverted to constrain the gas content in GCs during these epochs.  However, we should note that our study is also relevant for nuclear clusters hosting IMBHs, known to exist. In such galactic nuclei, multiple star-formation epoch is expected from infall of gas from larger scales. In our galaxy, $\Omega$-Cen is thought to be the nucleus of a past dwarf galaxy, and indeed shows evidence for clear multiple generations of star-formation epochs \citep{Woolley1966}.

    \item The dynamics of objects embedded in gas are complex and can be modeled using various approaches and techniques, which often yield differing results. Here, we employed disk migration (Section \ref{subsec:disk-migration}) and GDF (Section \ref{sec:gas_dynamical_friction}), but other models could potentially modify our findings.

    \item Our model necessitates the presence of an IMBH residing within a GC. Although this scenario is plausible and widely discussed in the literature, direct and conclusive evidence remains elusive. Future observations and theoretical advancements might challenge or even rule out this possibility.

    \item The lifetime and stability of the disk are influenced by a multitude of coupled and complex processes that involve the overall evolution of the cluster. These processes could lead to a shorter disk lifetime than our predictions. The disk itself could be disrupted by violent interactions with other stars.

    \item The coalescence of objects within the disk could trigger explosive events that might prematurely disrupt the disk. However, these events could also leave behind unique observational signatures.

    \item Global effects within the cluster could result in the segregation of stars and gas towards the center, potentially leading to feedback mechanisms that are not considered in our current model.

    \item Our primary focus was on gas-enriched clusters resulting from the gas associated with the second generation of star formation. As discussed earlier, other physical processes, such as accretion from the interstellar medium, could also enrich clusters with gas.

    \item Interactions with other stars could modify the inspiral, in the case of inspiral that takes longer than the relaxation time of the cluster, e.g, in the case of disk-migration dominated evolution.
    \item Winds or irradiation originating from the mAGN could affect the gas in the cluster, and potentially lead to its depletion or even unbinding. 
    
\end{itemize}

\section{Summary}\label{sec:summary}
This study explores the formation and dynamical implications of mini-active galactic nuclei (mAGNs) -- accretion disks around intermediate-mass black holes (IMBHs) within gas-rich globular clusters. We investigate conditions for stable mAGN disk formation and model the gas-induced migration of stars and black holes within them, emphasizing the dominant role of gas dynamical friction. 

The presence of mAGN disks could lead to several observable effects. These include alignment of stellar orbits, enhanced gravitational wave mergers with a potential frequency shift, unique tidal disruption events (mTDEs/cTDEs), and the formation of accreting IMBH systems possibly associated with ultraluminous X-ray sources. We estimate merger rates and the various types of TDE rates, highlighting potential observational signatures.

While our study reveals intriguing possibilities, we acknowledge uncertainties related to gas properties in young GCs, complex gas dynamics, and the lack of direct IMBH observations. Nevertheless, this work underscores the importance of mAGNs in understanding IMBHs, globular cluster evolution, and potentially the formation of supermassive black holes. Future multi-wavelength observations are crucial for confirming their existence and characterizing their properties.

\section*{Acknowledgements}

We would like to thank Evgeni Grishin for fruitful discussions, and Barry Ginat for fruitful discussions and a tea tray. 
MR acknowledges the generous support of Azrieli \& Rothschild fellowships. AAT acknowledges support from the Horizon Europe research and innovation programs under the Marie Sk\l{}odowska-Curie grant agreement no. 101103134.

%%%%%%%%%%%%%%%%%%%%%%%%%%%%%%%%%%%%%%%%%%%%%%%%%%
\section*{Data Availability}

The data underlying this article will be shared on reasonable request to the corresponding author.



\bibliographystyle{mnras}
\bibliography{example} % if your bibtex file is called example.bib

%\begin{thebibliography}{99}
%\end{thebibliography}

%%%%%%%%%%%%%%%%%%%%%%%%%%%%%%%%%%%%%%%%%%%%%%%%%%

%%%%%%%%%%%%%%%%% APPENDICES %%%%%%%%%%%%%%%%%%%%%

%%%%%%%%%%%%%%%%%%%%%%%%%%%%%%%%%%%%%%%%%%%%%%%%%%

% Don't change these lines
\bsp	% typesetting comment
\label{lastpage}
\end{document}